\documentclass[prb,twocolumn,superscriptaddress,showpacs, nofootinbib]{revtex4-2}
\usepackage{graphicx}
\usepackage{float}
\usepackage{dcolumn}
\usepackage{float}
\usepackage{bm}
\usepackage{mathrsfs}
\usepackage{ulem}
\usepackage[per-mode = fraction]{siunitx}
\usepackage{hyperref}
\hypersetup{
    colorlinks=true,
    linkcolor=blue,
    filecolor=magenta,      
    urlcolor=cyan,
    citecolor=magenta,
}
\usepackage{xcolor}
\usepackage{soul}
\usepackage{amsthm,amssymb}
\usepackage{mathtools}
\usepackage{comment}
\usepackage{physics}

\begin{document}

\title{Supercurrent rectification with time-reversal symmetry broken multiband superconductors}

\author{Yuriy Yerin}
\affiliation{CNR-SPIN, via del Fosso del Cavaliere, 100, 00133 Roma, Italy}
\author{Stefan-Ludwig Drechsler}
\affiliation{Institute for Theoretical Solid State Physics, Leibniz-Institut für Festkörper- und Werkstoffforschung IFW-Dresden, D-01169 Dresden, Helmholtzstraße 20}
\author{A. A. Varlamov}
\affiliation{CNR-SPIN, via del Fosso del Cavaliere, 100, 00133 Roma, Italy}
\affiliation{Istituto Lombardo ``Accademia di Scienze e Lettere'', via Borgonuovo,
25 - 20121 Milan, Italy}
\author{Mario Cuoco}
\affiliation{CNR-SPIN, c/o Universit\'a di Salerno, I-84084 Fisciano (SA), Italy}
\author{Francesco Giazotto}
\affiliation{NEST Istituto Nanoscienze-CNR and Scuola Normale Superiore, I-56127, Pisa, Italy}

\date{\today}

\begin{abstract}
\textcolor{black}{We consider nonreciprocal supercurrent effects in Josephson junctions based on multiband superconductors with a pairing structure that can break time-reversal symmetry. We demonstrate that a nonreciprocal supercurrent can be generally achieved by the cooperation of interband superconducting phase mismatch and interband scattering as well as by multiband phase frustration. The effect of interband impurity scattering indicates that the amplitude and sign of the nonreciprocal supercurrent are sensitive to the interband phase relation. 
For the case of a three-band superconductor, due to phase frustration, we show that the profile of the supercurrent rectification is marked by a hexagonal pattern of nodal lines with vanishing amplitude. 
Remarkably, around the nodal lines, the supercurrent rectification amplitude exhibits three-fold structures with an alternating sign. We show that the hexagonal pattern and the three-fold structure in the interband phase space turn out to be dependent on the tunneling amplitude of each band. These findings provide hallmarks of the supercurrent rectification which can be potentially employed to unveil the occurrence of spin-singlet multiband superconductivity with time-reversal symmetry breaking.}
\end{abstract}
\maketitle

\section{Introduction}

Nonreciprocal effects in superconductors are generally based on phenomena where the amplitude of the supercurrent depends on the direction of its flow.
The rectification of the supercurrent is then a direct result of the nonreciprocal superconducting transport.
Recently, amount of works has been devoted to the achievement of supercurrent rectification \cite{and20,baur22, bau22, wu22,jeo22,nad23,Ghosh2024} both motivated by fundamental questions on the underlying generating mechanisms 
as well as on the technological challenges regarding the development of dissipationless electronics and quantum circuits \cite{Upadhyay2024}. 
Nonreciprocal supercurrent phenomena, apart from conventional superconductors, indeed, have been successfully demonstrated in several materials including non-centrosymmetric superconductors \cite{WakatsukiSciAdv2017,and20,Zhang2020}, two-dimensional electron gases and polar semiconductors\cite{Itahashi20}, patterned superconductors\cite{Lyu2021}, superconductor-magnet hybrids \cite{wu22,Narita2022}, Josephson junctions with magnetic atoms \cite{tra23}, twisted graphene systems \cite{lin22}, and high T-$_c$ superconductors \cite{Ghosh2024}.

Several mechanisms, either intrinsic or extrinsic in nature, have been proposed to devise a supercurrent diode with control of the rectification in sign and amplitude. 
In this context it is generally accepted that breaking of time-reversal and inversion symmetries are key requirements for achieving a nonreciprocal supercurrent. For instance, Cooper pair momenta \cite{lin22,pal22,yuan22} or helical phases \cite{Edelstein95,Ilic22,dai22,he22,Turini22}, as well as screening currents \cite{hou23,sun23}, and supercurrents related to self-induced field \cite{kras97,GolodNatComms2022} have been considered as physical scenarios and mechanisms to get nonreciprocal supercurrent effects.
The breaking of time-reversal symmetry is mostly achieved through external magnetic fields. Vortices are also expected to yield supercurrent diode effects and their role has been investigated for a variety of physical configurations \cite{GolodNatComms2022,sur22,GutfreundNatComms2023,Gillijns07,Ji21,He19,MarginedaCommunPhys2023,Paolucci23,Greco23,Lustikova2018,Itahashi20}. Proposals of magnetic field-free superconducting diodes have been using magnetic materials in suitably designed heterostructures \cite{wu22,Narita2022}. Instead, back-action supercurrent mechanisms and electrical gating \cite{margineda2023backaction} can result into superconducting rectification effects without the use of external magnetic fields or magnetic materials. 

In this context, the exploitation of an unconventional superconducting state with intrinsic time-reversal symmetry breaking represents a potential path to get supercurrent rectification. This is indeed an open problem which has not been fully explored yet.
The role of symmetry-broken phases in the normal state and the occurrence of time-reversal broken superconducting phases have been addressed only for two-dimensional materials demonstrating that indeed they can lead to a distinct type of supercurrent rectification without the need of external magnetic fields \cite{Scammell_2022}.

Several superconducting materials exhibit signatures of spontaneous time-reversal symmetry breaking (TRSB) below the transition temperature \cite{Kallin_2016,Wyso2019,Ghosh_2021}.
In this type of superconductors the occurrence of internal magnetic fields is either due to the magnetic moment of the Cooper pairs, as in non-unitary spin-triplet pairing, or by the multicomponent nature of the superconducting condensate in multiband superconductors. 
For the latter, it is the complex superposition of distinct order parameters that leads to a breaking of the time-reversal symmetry. In this context, it is known that disorder can cooperate to favor the formation of time-reversal symmetry broken phases both in multiband superconductors \cite{Stanev,Corticelli,Lee2009,Maiti2015}. In particular, for spin-singlet multiband superconductors a distinct role is played by the so-called $\pi$ pairing, i.e., the antiphase relation between the order parameters in different bands, or equivalently, the sign reversal of the Josephson coupling between Cooper pairs in different bands. Apart from the connection with the time-reversal symmetry breaking, the intertwinning of $\pi$ pairing and multiband electronic structure often marks the occurrence of unconventional superconducting phases, e.g. in iron-based \cite{Grinenko2020,Grinenko2021} and oxide interface superconductors \cite{Scheurer2015,Singh2022}, electrically or orbitally driven superconductivity \cite{Mercaldo2020,Bours2020,Mercaldo2021,DeSimoni2021,Mercaldo2023}, and multiband noncentrosymmetric superconductors \cite{Scheurer2015,Fukaya2018,Fukaya2020,fukaya22}. 
Understanding the mechanisms for time-reversal symmetry broken multiband phases as well as identifying specific
detection schemes for accessing the complexity of multiband superconductors are key challenges not yet fully settled.

In this paper, we consider multiband superconductors that can break the time-reversal symmetry due to nontrivial interband phase relation and we study the character of nonreciprocal supercurrents effects.  
We demonstrate that nonreciprocal supercurrent in Josephson junctions with multiband superconductors can be generally achieved by exploiting the combination of the interband superconducting phase mismatch and the strength of the interband scattering. 
In particular, we find that the effect of interband impurity scattering can help to distinguish among a dominant 0 or $\pi$ pairing in the superconductor. We show that distinct variations in the amplitude and sign of the rectification amplitude can be observed as a function of the interband impurity scattering strength. 
In the case of three-band superconductors there is a phase frustration in the complex superposition of the order parameters that results in the formation of nodal lines or large regions in the parameters space with vanishing rectification amplitude. The nodal patterns are marked by multifold structures with sign changes of the rectification amplitude. The location and shape of these structures depend on the tunneling amplitudes. 

The paper is organized as follows. In  Sect. II we present supercurrent nonreciprocal effects in a Josephson junction combining single-band and two-band superconductors with time-reversal symmetry-breaking pairing by focusing on the role of interband impurity scattering. Section III is devoted to a multiband phase frustrated configuration. There, we present the study of a Josephson junction hosting a three-band superconductor interfaced with a convectional single-band superconductor. The conclusions are given in Sect. IV.  

\section{The Josephson current between a single-band and a two-band superconductor}

In this Section, we demonstrate how the interplay of interband impurity scattering and nontrivial interband phase can yield supercurrent nonreciprocal effects. This is done by considering a Josephson junction composed of a conventional single-band superconductor interfaced to a superconductor with two-bands whereas the time-reversal symmetry breaking arises from the nontrivial interband superconducting phase relation. 

Let us start with the description of the s-wave two-band superconductor in the presence of impurity scattering and zero magnetic field. In the dirty limit the Eilenberger formalism for a two-band superconductor, as for the conventional single-band s-wave counterpart, can be reduced to the Usadel equations for quasiclassical Green's functions decomposed in spherical harmonics \cite{Gurevich}:
\begin{equation}
\label{Usadel_2band}
\omega {f_i}\! -\! {D_i}\left( {g_i}{\nabla ^2{f_i} \!- \!{f_i}{\nabla ^2}{g_i}} \right) \!=\! {\Delta _i}{g_i} \!+ \!{\Gamma _{ij}}\left( {{g_i}{f_j}\! -\! {g_i}{f_j}} \right),
\end{equation}
where $f_i=f_i({\bf{r}},\omega)$, $g_i=g_i({\bf{r}},\omega)$ are the $\bf{r}$ coordinate-dependent anomalous and normal quasiclassical Green's functions connected by the standard normalization condition ${g_i}^2 + {\left| {{f_i}} \right|^2} = 1$,  $\{i, j\} = 1,2$. The remaining notations are: $\omega \equiv \omega_n = \left( {2n + 1} \right)\pi T$ is the Matsubara frequency, ${D_i}$ are the intraband diffusion coefficients caused by the intraband elastic scattering, $\Delta_i$ represent complex order parameters in a two-band superconductor. Finally, $\Gamma _{ij}$ denote the interband scattering coefficients, which are absent in the case of a clean multi-band superconductor. When $\Gamma_{ij}=0$ Eq. (\ref{Usadel_2band}) can be decoupled and the Green's functions of different bands are related only through the interband interaction in the self-consistency equations for the order parameters.

To determine the supercurrent, Eq. (\ref{Usadel_2band}) must be evaluated together with the expression for the current density
\begin{equation}
\label{current_general}
j({\bf{r}}) =  - e\pi {\rm i} T\sum\limits_i {\sum\limits_{\omega }^{ } {{N_i}{D_i}\left( {f_i^*\nabla {f_i} - {f_i}\nabla f_i^*} \right)} },
\end{equation}
where $N_i$ corresponds to the partial contribution of each band to the density of states at the Fermi level.

Now, we proceed with the study of the properties of the Josephson junction between the standard single-band superconductor (left lead) and the multi-band one (right lead). The Josephson junction is studied as a weak link connecting two superconducting leads in the form of a thin short filament of length $L$.  Based on the condition that the length $L \lesssim \min {\xi _i}\left( 0 \right)$ [${\xi _i}\left( T \right)$ being the temperature-dependent coherence lengths] one can consider Josephson junction as a quasi-one-dimensional system and neglect all terms in Eq. (\ref{Usadel_2band}) except the gradient one \cite{Kulik1}
\begin{equation}
\label{Usadel_short}
g_i \frac{{{d^2}}}{{d{x^2}}}{f_i} - {f_i}\frac{{{d^2}}}{{d{x^2}}} g_i  = 0.
\end{equation}

Eq. (\ref{Usadel_short}) can be solved using the parameterized functions $\Phi_i$ which are connected with the Green's functions $f_i$ and $g_i$ by the following expressions (see e.g. Ref. \onlinecite{Golubov_review})
\begin{equation}
\label{Phi_intro}
{g_i} = \frac{\omega }{{\sqrt {{\omega ^2} + {\Phi _i}\Phi _i^ * } }},{f_i} = \frac{{{\Phi _i}}}{{\sqrt {{\omega ^2} + {\Phi _i}\Phi _i^ * } }},{\Phi _i} = \frac{{\omega {f_i}}}{{{g_i}}}.
\end{equation}
With this parameterization the normalization conditions for the Green functions are fulfilled automatically. We notice that in the single-band superconducting lead of the Josephson junction at $x=-L/2$ the function $\Phi_0=\Delta_0$. Here and hereafter the zero subscript is attributed to the parameters of the left-side of the Josephson junction. Furthermore, we suppose that the critical temperature $T^{(s)}_{c}$ of the bulk single-band superconducting bank of the Josephson junction is at least not less than that of the bulk multi-band superconductor, i.e., it will be assumed that the $T^{(s)}_{c} \geq T^{(m)}_{c}$.

The Josephson phase $\varphi$, which we define as the difference between the phases of the order parameter  $\Delta_0$ of the single-band superconductor (left lead) and the order parameter $\Delta_1$ of the multiband superconductor (right lead), determines the boundary condition
\begin{equation}
{\Delta _0} (-L/2) = \left| {{\Delta _0}} \right|\exp \left( { - {\rm i}\varphi/2 }\right). 
\end{equation}

Now let us determine the boundary conditions for the Green's functions in Eqs. (\ref{Usadel_short}) for the two-band superconducting bank (at $x= L/2$). They can be found in the homogeneous expressions of Eqs. (\ref{Usadel_2band}) where the gradient terms are now being ignored. In general case, there is no analytical solution for $\Gamma_{ij} \ne 0$. However, close to $T^{(m)}_{c}$ an amenable conjecture for the Green's functions can be found by the method of successive approximations over the moduli of the order parameters $\Delta_i$, which yields \cite{Stanev}
\begin{equation}
\label{f_i0}
\begin{array}{l}
f_1^{\left( 0 \right)} = \frac{{\left( {\omega  + {\Gamma _{21}}} \right){\Delta _1} + {\Gamma _{12}}{\Delta _2}}}{{\omega \left( {\omega  + {\Gamma _{12}} + {\Gamma _{21}}} \right)}},\\
f_2^{\left( 0 \right)} = \frac{{\left( {\omega  + {\Gamma _{12}}} \right){\Delta _2} + {\Gamma _{21}}{\Delta _1}}}{{\omega \left( {\omega  + {\Gamma _{12}} + {\Gamma _{21}}} \right)}},
\end{array}
\end{equation}
and
\begin{widetext}
\begin{equation}
\label{f_i1}
\begin{array}{l}
f_1^{\left( 1 \right)} = \frac{{{\Gamma _{12}}\left( {\omega  + {\Gamma _{21}}} \right)\left( {{\Delta _1} - {\Delta _2}} \right){{\left| {f_2^{\left( 0 \right)}} \right|}^2} - \left[ {\left( {{{\left( {\omega  + {\Gamma _{21}}} \right)}^2} + {\Gamma _{12}}\left( {\omega  + 2{\Gamma _{21}}} \right)} \right){\Delta _1} + {\Gamma _{12}}\left( {\omega  + {\Gamma _{12}}} \right){\Delta _2}} \right]{{\left| {f_1^{\left( 0 \right)}} \right|}^2}}}{{\omega \left( {\omega  + {\Gamma _{12}} + {\Gamma _{21}}} \right)}},\\
f_2^{\left( 1 \right)} = \frac{{{\Gamma _{21}}\left( {\omega  + {\Gamma _{12}}} \right)\left( {{\Delta _2} - {\Delta _1}} \right){{\left| {f_1^{\left( 0 \right)}} \right|}^2} - \left[ {\left( {{{\left( {\omega  + {\Gamma _{12}}} \right)}^2} + {\Gamma _{21}}\left( {\omega  + 2{\Gamma _{12}}} \right)} \right){\Delta _2} + {\Gamma _{21}}\left( {\omega  + {\Gamma _{21}}} \right){\Delta _1}} \right]{{\left| {f_2^{\left( 0 \right)}} \right|}^2}}}{{\omega \left( {\omega  + {\Gamma _{12}} + {\Gamma _{21}}} \right)}}.
\end{array}
\end{equation}
\end{widetext}
In Eqs. (\ref{f_i0})- (\ref{f_i1}) the values of the order parameters on the right side of the junction are
\begin{equation}
{\Delta _1} (L/2) = \left| {{\Delta _1}} \right|\exp \left( { {\rm i}\varphi/2 }\right),
\end{equation}
and
\begin{equation}
{\Delta _2}(L/2) = \left| {{\Delta _2}} \right|\exp \left( { {\rm i}\varphi/2 + {\rm i}\phi }\right).
\end{equation} 
Here $\phi$ is the intrinsic difference between the phases of the order parameters of the two-band superconductor that can attain non-zero values due to the presence of the interband scattering ($\Gamma_{ij} \neq 0$) solely \cite{Stanev, Corticelli, Babaev_PD, Yerin_magneto}. In the absence of the latter, the values of $\phi$ can be equal to $0$ or $\pi$ depending on the attractive or repulsive nature of the interband interaction and corresponds to $s_{++}$ or $s_{\pm}$ pairing symmetry, respectively. Otherwise, for $\phi$ different from $0$ or $\pi$, a two-band superconductor is characterized by a complex superposition of the two order parameters that thus breaks the time-reversal symmetry. 

We notice that the solutions for the anomalous Green's functions 
cannot be used for the entire temperature range of superconducting state, since the applicability of Eqs. (\ref{f_i0})- (\ref{f_i1}) is restricted by the region where both $\left| {{\Delta _i}} \right|$ are small, {\it i.e.} nearby the critical temperature $T^{(m)}_{c}$. In turn, the latter depends not only on the intra- and interband interactions but also on the interband scattering rate $\Gamma=\Gamma_{12}=\Gamma_{21}$ (with the additional assumption $N_1=N_2$), which lowers the value of $T^{(m)}_{c}$ as compared to the critical temperature $T_{c0}$ of a clean two-band superconductor \cite{Gurevich, Stanev}. Correspondingly, our study of the current-phase relations will be performed for temperatures close enough to the $T^{(m)}_{c}$ of the dirty two-band superconductor. 

\begin{figure}
\includegraphics[width=0.49\columnwidth]{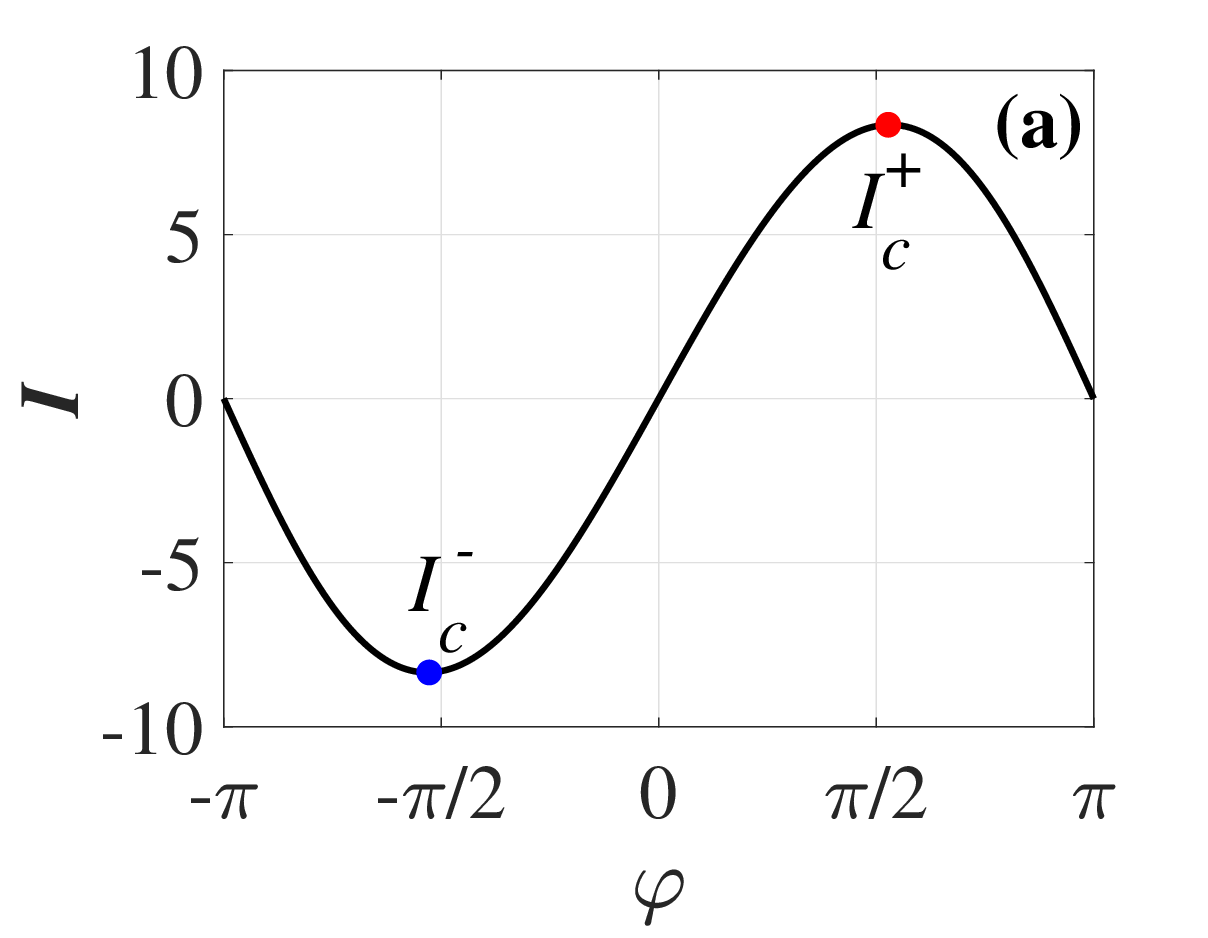}
\includegraphics[width=0.49\columnwidth]{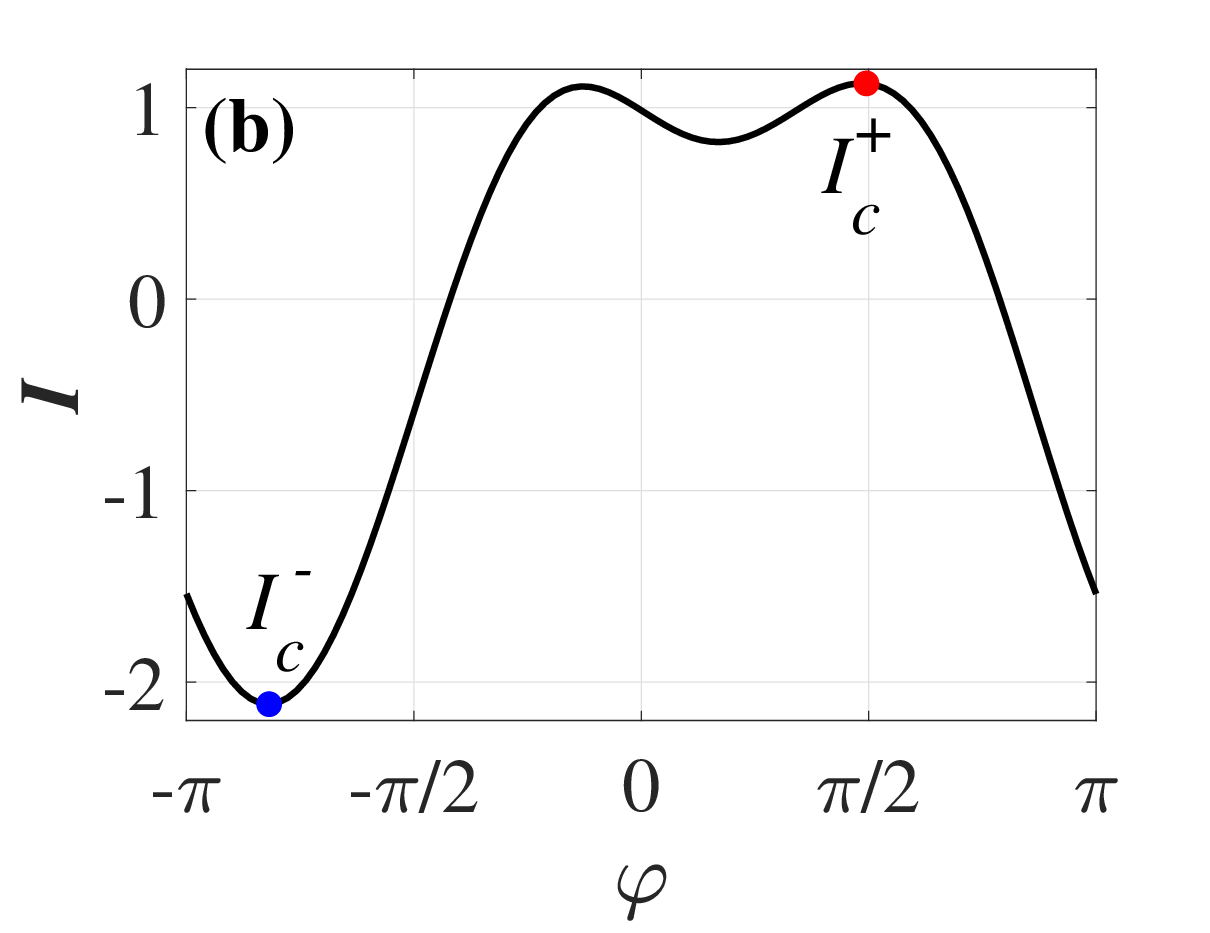}
\caption {Current-phase relation for the Josephson junction between a single-band and an $s_{++}$ two-band superconductor with (a) $\Gamma= 0$ and $\phi=0$ and an $s_{\pm}+is_{++}$ two-band superconductor with representative values of the interband scattering,
$\Gamma  \approx 0.062 T_{c0}$, and the interband phase difference, $\phi \approx \frac{{3}}{{25}}\pi$ (b). We employ representative values of the order parameters in the two-band superconductors:  $\left| {{\Delta _1}} \right| = 2\left| {{\Delta _0}} \right|$, $\left| {{\Delta _2}} \right| = 3\left| {{\Delta _0}} \right|$ at $T=0.7\,T_{c0}$. The amplitude of the supercurrent $I$ is taken in units of $\frac{{\pi \left| \Delta_0  \right|}}{{e{R_{N1}}}}$.}
\label{CPR_2band}
\end{figure}

Eqs. (\ref{Usadel_short}) with boundary conditions Eqs. (\ref{f_i0}) and (\ref{f_i1}) admit an analytical solution. Substituting the solutions of Eqs. (\ref{Usadel_short}) to Eq. (\ref{current_general}) for the current density we derive the explicit expression for the Josephson current flowing between the single-band and the dirty two-band superconductor:
\begin{widetext}
\begin{equation}
\label{current_total}
I(\varphi) = \frac{{\pi T}}{e}\sum\limits_i {\frac{1}{{{R_{Ni}}}}} \sum\limits_\omega  {\frac{{{C_i}}}{{\sqrt {1 - \kappa _i^2 + C_i^2} }}} \left( {\arctan \left( {\frac{{\omega {\kappa _i}{C_i} + \Phi _i^ - \left( {{\kappa ^2} - 1} \right)}}{{\omega \sqrt {\kappa _i^2 - C_i^2 - 1} }}} \right) - \arctan \left( {\frac{{\omega {\kappa _i}{C_i}{\mkern 1mu}  + \Phi _0^ - \left( {{\kappa ^2} - 1} \right){\mkern 1mu} }}{{\omega \sqrt {\kappa _i^2 - C_i^2 - 1} }}} \right)} \right).
\end{equation}
\end{widetext}
Here the Josephson phase difference $\varphi$ enters implicitly via the functions
\begin{equation}
\label{kappa_i}
{\kappa _i} = \frac{{\Phi _0^ + {\mkern 1mu}  - \Phi _i^ + }}{{\Phi _0^ - {\mkern 1mu}  - \Phi _i^ - }},
\end{equation}
\begin{equation}
\label{C_i}
{C_i} = \frac{{\Phi _0^ - {\mkern 1mu} \Phi _i^ +  - \Phi _0^ + {\mkern 1mu} \Phi _i^ - }}{{\omega \left( {\Phi _0^ -  - \Phi _i^ - } \right)}},
\end{equation}
where
\begin{equation}
\begin{array}{l}
\Phi _0^ \pm  = \frac{1}{2}\left( {{\Phi _0} \pm \Phi _0^ * } \right),\\
\Phi _i^ \pm  = \frac{1}{2}\left( {{\Phi _i} \pm \Phi _i^ * } \right),
\end{array}
\end{equation}
and $R_{Ni}$ are partial contributions to the junction's resistance (also referred as Sharvin resistance for the case of point contacts \cite{Sharvin}). In the following, we assume $R_{N1}=R_{N2}=R_{N3}$. 

For the case of Josephson junction between two different s-wave single-band superconductors, Eq. (\ref{current_total}) yields the current-phase relation already obtained in Ref. \cite{Zubkov}. Another important remark is that the derivation of Eq. (\ref{current_total}) was done in the single-channel limit for a disordered regime of diffusive type for the Josephson junction. 

Using Eq. (\ref{current_total}) one can plot the current-phase relation $I(\varphi)$ of the Josephson junction between a single-band superconductor and a two-band superconductor with the $s_{++}$ (Fig. \ref{CPR_2band}a) and the chiral order parameter pairing, i.e. $s_{+-}+i s_{++}$ (see Fig. \ref{CPR_2band}b). For the sake of clarity, in these figures we introduce the notation of the critical current $I_c^+$ (filled red dot) for the maximum forward supercurrent 
and $I_c^-$ (filled blue dot) for the maximum negative amplitude of the supercurrent. 

As one can see from Figure  \ref{CPR_2band}a in the case of $s_{++}$ pairing symmetry, when $\phi=0$ and $\Gamma=0$, the current phase relation is symmetric and the supercurrent exhibits a reciprocal behavior \cite{Yerin_two_band, Yerin_review}. As expected, this behavior is qualitatively consistent 
with that one of the Josephson junction between two different single-band s-wave superconductors separated by a very thin normal layer \cite{Zubkov} or to the that of standard Josephson junction with constriction (S-c-S type) \cite{Kulik1, Kulik2}. For this pairing state the forward ($I_c^+$ ) and backward ($I_c^-$)  critical current of the junction are identical, i.e. $I_c^+=I_c^-$.  

The current-phase relation is substantially different when considering the effect of interband scattering assuming a complex superposition of 0  and $\pi$ pairing order parameters (i.e. $s_{\pm}+is_{++}$). 
First of all, as one can see from Figure \ref{CPR_2band}b that the current phase relation $I(\varphi)$ becomes non-sinusoidal, in contrast to the similar pattern in Figure \ref{CPR_2band}a. Moreover, it has the asymmetry of the so-called $\varphi_0$ Josephson junction, i.e. at $\varphi = 0$ the supercurrent amplitude is not zero \cite{Buzdin, Buzdin_review}. Another remarkable feature of this current-phase relation is that for specific values of $\Gamma$ and the intrinsic phase difference $\phi$, the critical currents $I_c^+$ and $I_c^-$ of the Josephson junction can be substantially different in amplitude, with $I_c^+$ which can be even become identically zero. 

\begin{figure}[h]
\includegraphics[width=0.98\columnwidth]{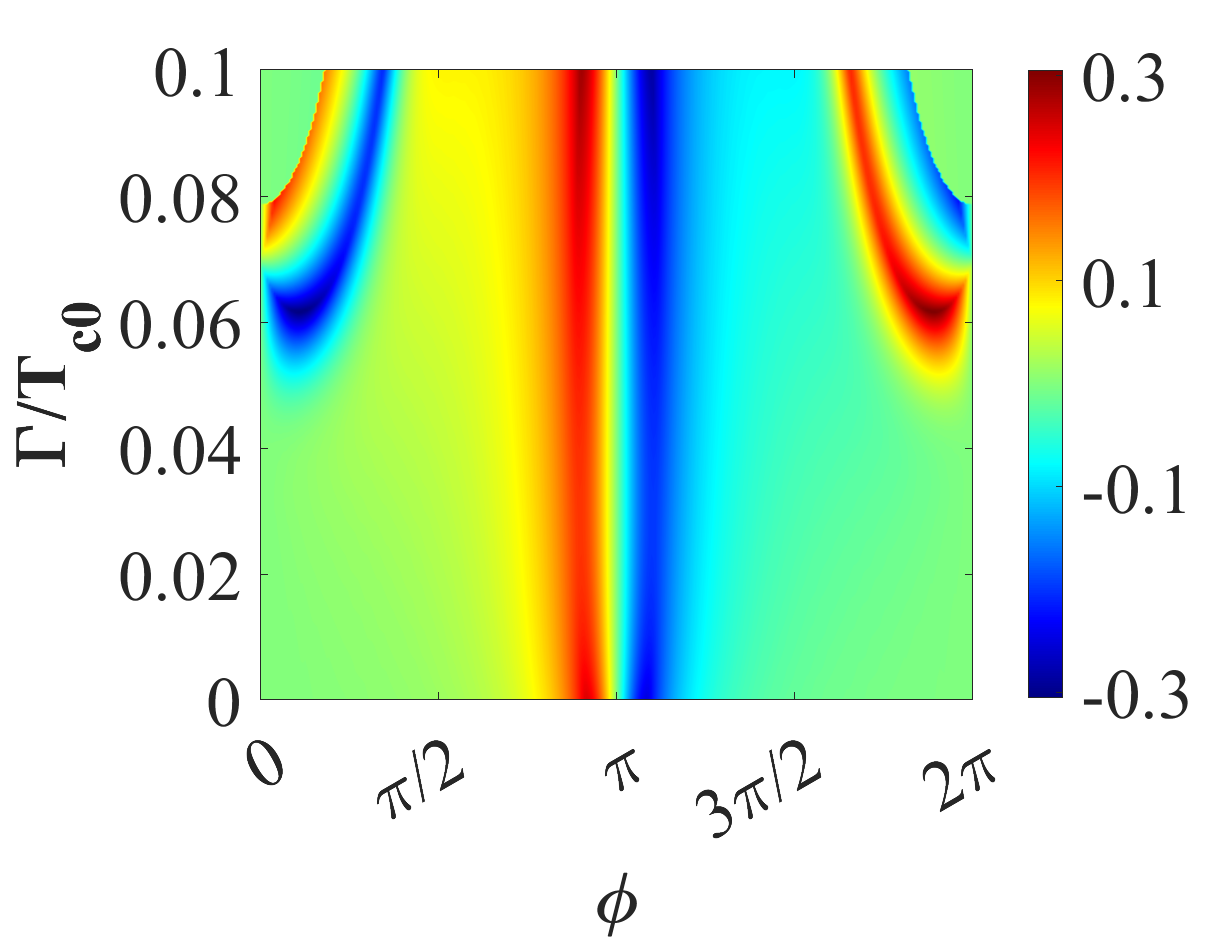}
\caption {Contour map of the supercurrent rectification amplitude, $\eta$, for the Josephson junction as a function of the phase difference $\phi$ and the interband scattering rate $\Gamma/T_{c0}$ assuming a two-band superconductor with 
$\left| {{\Delta _1}} \right| = 2\left| {{\Delta _0}} \right|$, $\left| {{\Delta _2}} \right| = 3\left| {{\Delta _0}} \right|$ at $T=0.7T_{c0}$. $\eta$ is determined by evaluating the difference among the maximal amplitude of the critical currents for forward ($I_c^+$) and backward ($I_c^-$) flow directions as in Eq. 14.}
\label{Diod_2band}
\end{figure}

Such an unusual asymmetric pattern of the current-phase relation in Josephson junction hosting a single-band and a two-band superconductor opens the path for achieving nonreciprocal supercurrent effects guided by multiband time-reversal symmetry broken phases.
The corresponding diode rectification amplitude, $\eta$, can be determined by evaluating the difference among the maximal amplitudes of the critical currents for forward ($I_c^+$) and backward ($I_c^-$) flow directions: 
\begin{equation}
\label{diode_eff}
\eta  = \frac{{I_c^ + {\mkern 1mu}  - \left| {I_c^ - } \right|}}{{I_c^ + {\mkern 1mu}  + \left| {I_c^ - } \right|}} \,.
\end{equation}

We start by observing that the rectification amplitude for the current-phase relation as shown in Figure \ref{CPR_2band}a is $\eta=0$ ($I_c^+=|I_c^-|$) while that one corresponding to Figure \ref{CPR_2band}b can reach the value $\eta \approx -0.3$.
One can then track the evolution of the current phase relation and extract the corresponding rectification amplitudes for any value of the interband phase difference $\phi$ and interband scattering rate $\Gamma$.
The outcome of this analysis is reported in Fig. \ref{Diod_2band}.
We find that the rectification amplitude has a distinct dependence on the interband scattering rate when comparing the pairing configuration at $\phi < \pi/4$ with that close to $\phi \sim \pi$.
Indeed, in the former region, close to 0-interband pairing coupling, the rectification amplitude of the supercurrent is negligible below a critical threshold for the interband scattering rate $\Gamma$. Then, the increase of $\Gamma$ leads to a growth of the rectification amplitude and a subsequent sign reversal with equally sized rectification states.
By inspection of the current phase relation across the transition, one can grasp the origin of the achieved rectification. This is due to the occurrence of a 0-to-$\pi$ Josephson phase transition that is marked by a sign change of the odd-parity first harmonic component. It is known that in general, the supercurrent rectification value is sizable when the first and second harmonics are comparable in amplitude \cite{fukaya2024}. For this reason, since the second harmonic component is typically smaller than the first harmonic when approaching the 0-to-$\pi$ phase transition we have that the first and second harmonics get similar amplitude.


We point out that, in principle, for the constructed phase diagram of the superconducting diode rectification amplitude, of course, only the states with a phase difference $\phi$ different from zero in a two-band superconductor can be realized only when $\Gamma \ne 0$. Therefore, the lower part of the diagram adjacent to the $\phi$-axis are limiting cases that cannot be physically achieved when $\Gamma$ is vanishing. 
However, as can be seen in Fig. \ref{Diod_2band}, the highest rectification amplitude $\eta$ is attained away from the above-mentioned region of the phase diagram. Moreover, there can be other mechanisms to get a nontrivial phase difference $\phi$ due to intra- and interband interaction (see for instance Eq. (7) in Ref. \onlinecite{Stanev} or Eq. (9) in Ref. \onlinecite{Yerin_magneto}). 

\section{The Josephson current between a single-band and a three-band superconductor}

The replacement of the two-band superconductor by a three-band counterpart in the Josephson junction leads to additional phase interference effects and, as a consequence, yields a different structure in the current-phase relation \cite{Huang, Yerin_3band, Kiyko, Yerin_review, Xu} as well as frustration effects in the dynamics \cite{Guarcello2022}. Since the Josephson current is now determined by three partial contributions instead of two, for convenience and clarity, we exclude the effect of interband scattering and put $\Gamma_{ij}=0$, i.e. we consider a clean three-band superconductor.

\begin{figure}[h]
\includegraphics[width=0.49\columnwidth]{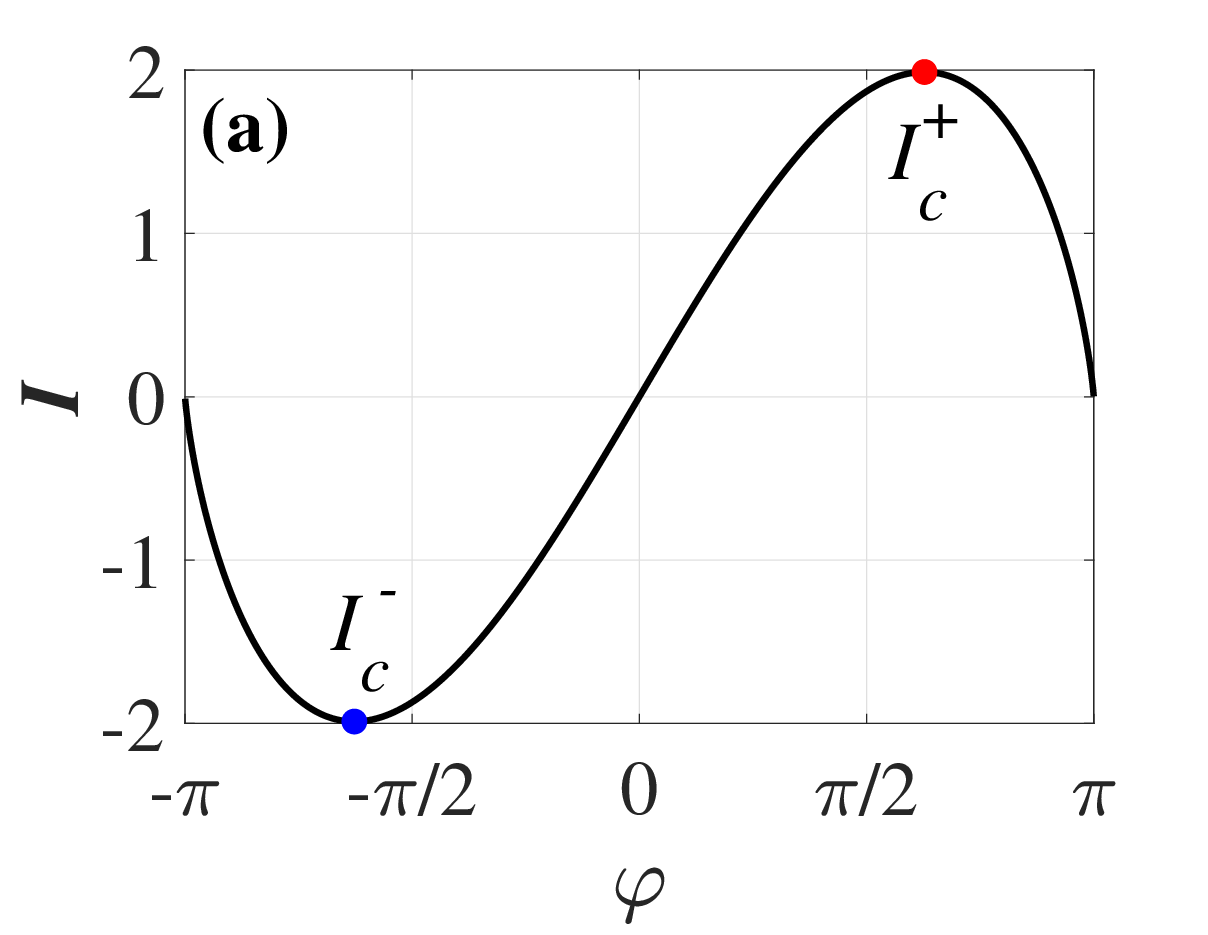}
\includegraphics[width=0.49\columnwidth]{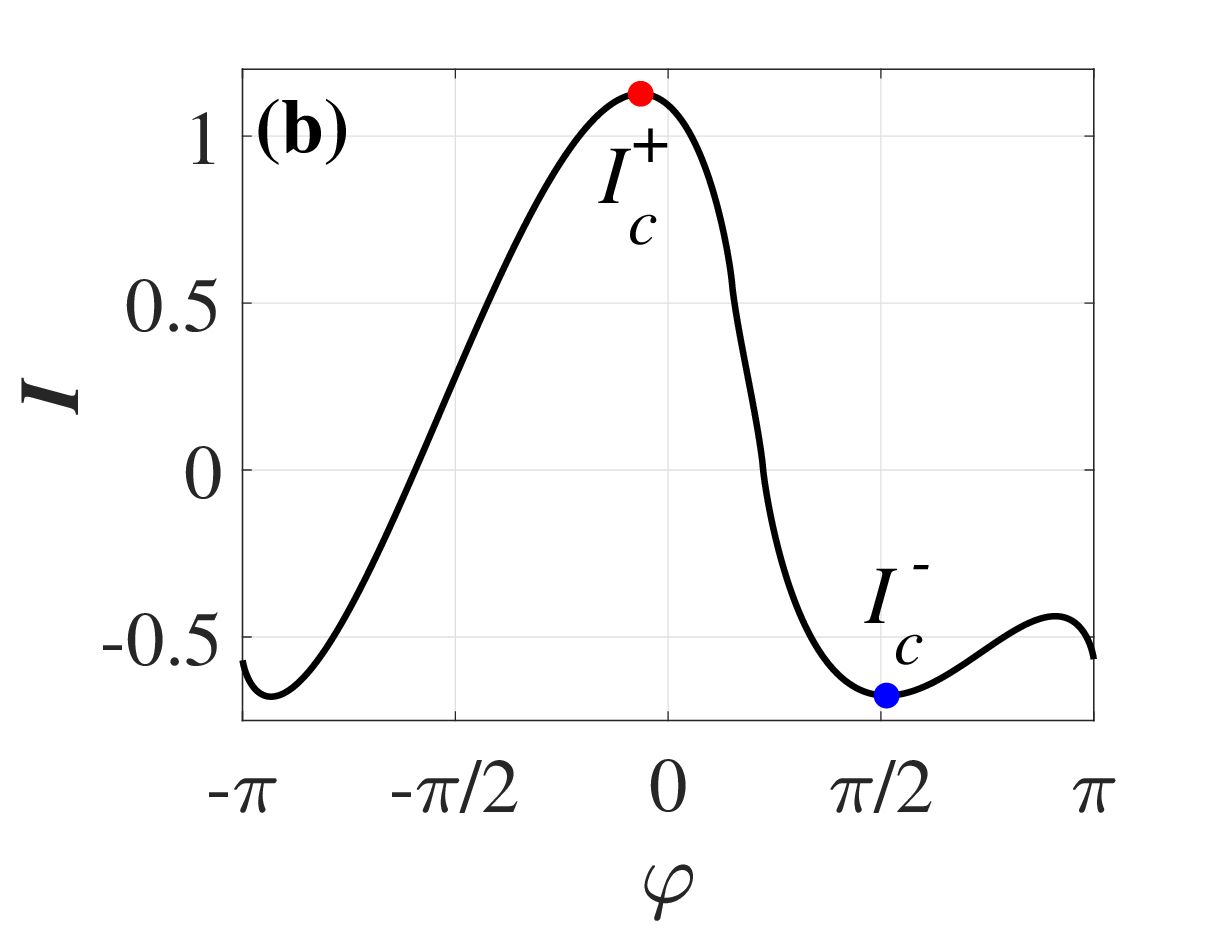}
\caption {Current-phase relations for the Josephson junction between single-band and three-band superconductor with $\phi=\theta=0$ (a) and 
$\phi \approx 1.22$, $\theta \approx 1.33$ (in radians)
(b) for $\left| {{\Delta _1}} \right| = \left| {{\Delta _2}} \right| = \left| {{\Delta _3}} \right| = \left| {{\Delta _0}} \right|$ at $T=0$. Current is measured in units of $\frac{{\pi \left| \Delta_0  \right|}}{{e{R_{N1}}}}$.}
\label{CPR_3band}
\end{figure}

Eq. (\ref{current_total}) remains applicable for the given Josephson system with the only correction now for the contribution to the current from the third band of the three-band superconductor.
Based on the assumption $\Gamma_{ij}=0$ one can find an exact expression for the boundary conditions without the necessity of solving the Eqs. (\ref{Usadel_2band}) by the method of successive approximations like in a two-band case. Here, boundary conditions acquire the form 
\begin{equation}
{\Phi_0} (-L/2) = \left| {{\Delta _0}} \right|\exp \left( -{\rm i}\varphi /2 \right),
\end{equation}
for the left lead (single-band superconductor) and 
\begin{eqnarray}
{\Phi_1} (L/2) &=& \left| {{\Delta _1}} \right|\exp \left( { {\rm i}\varphi/2 }\right), \nonumber \\
{\Phi _2}(L/2) &=& \left| {{\Delta _2}} \right|\exp \left( { {\rm i}\varphi/2 + {\rm i}\phi }\right), \nonumber \\
{\Phi _3}(L/2) &=& \left| {{\Delta _3}} \right|\exp \left( { {\rm i}\varphi/2 + {\rm i}\theta }\right),
\end{eqnarray}
for the right bank (three-band superconductor). The intrinsic phase differences $\phi$ and $\theta$ correspond to the pairs of order parameters $\Delta_1, \Delta_2$ and $\Delta_1, \Delta_3$ of a bulk three-band superconductor \cite{Brink}.

\begin{figure}[h!]
\includegraphics[width=0.89\columnwidth]{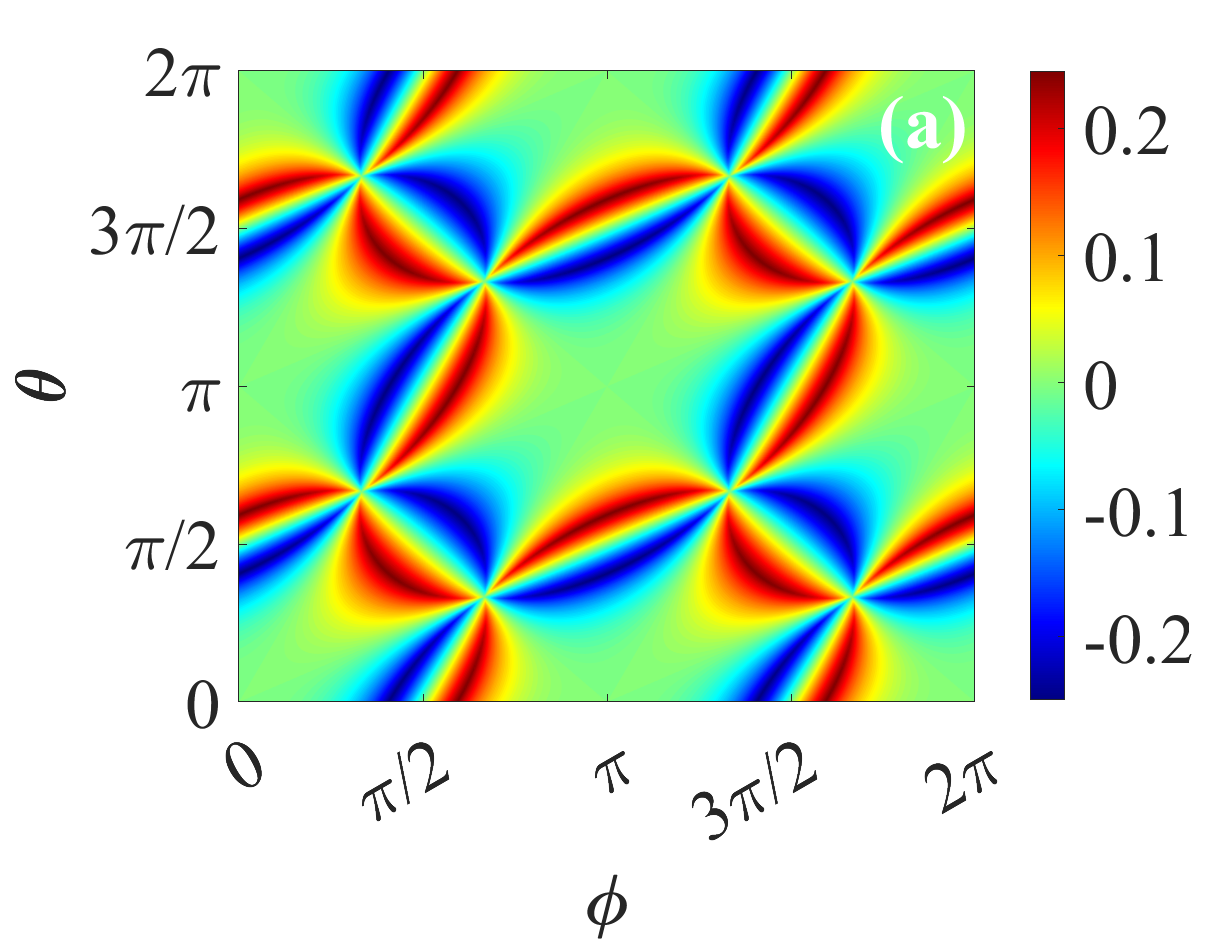}
\includegraphics[width=0.89\columnwidth]{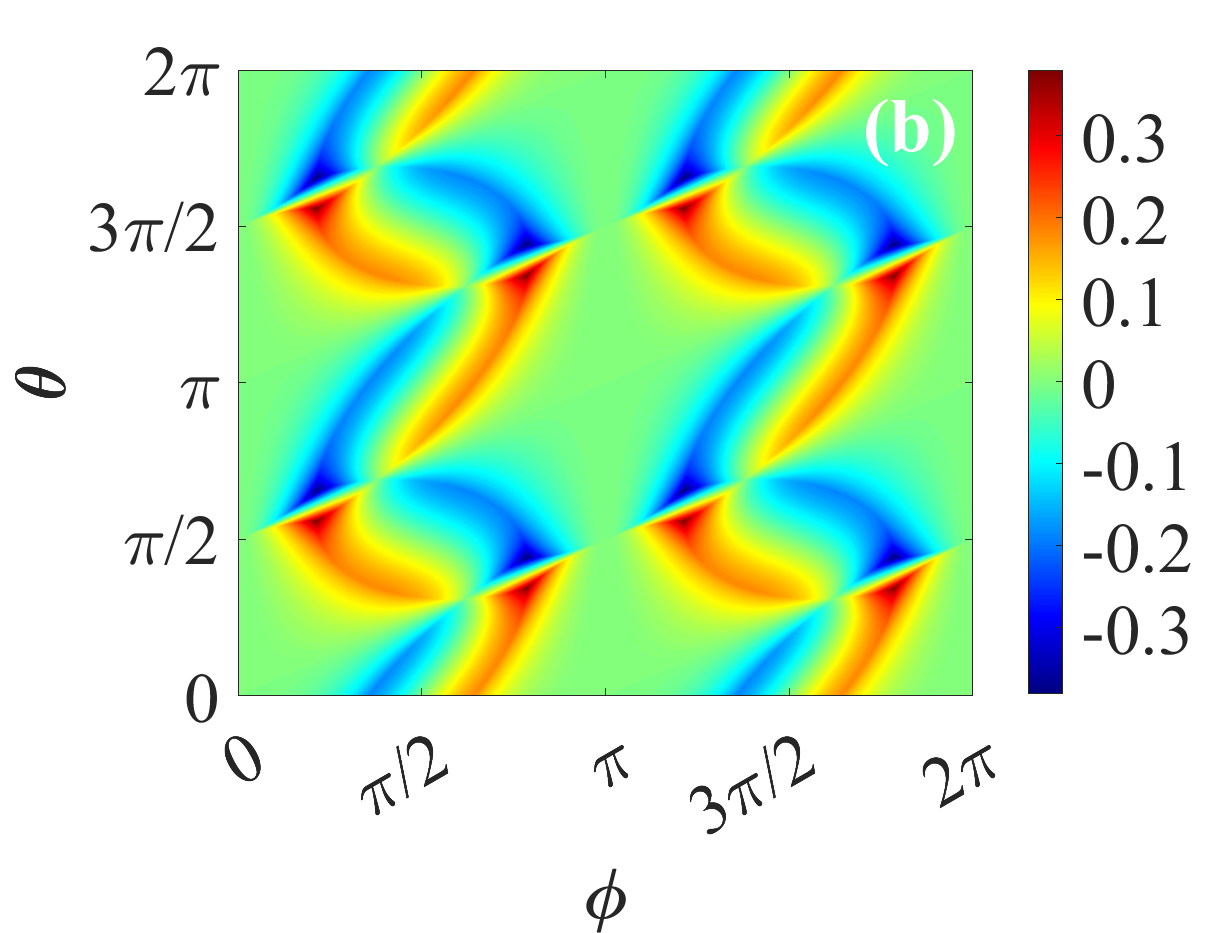}
\includegraphics[width=0.89\columnwidth]{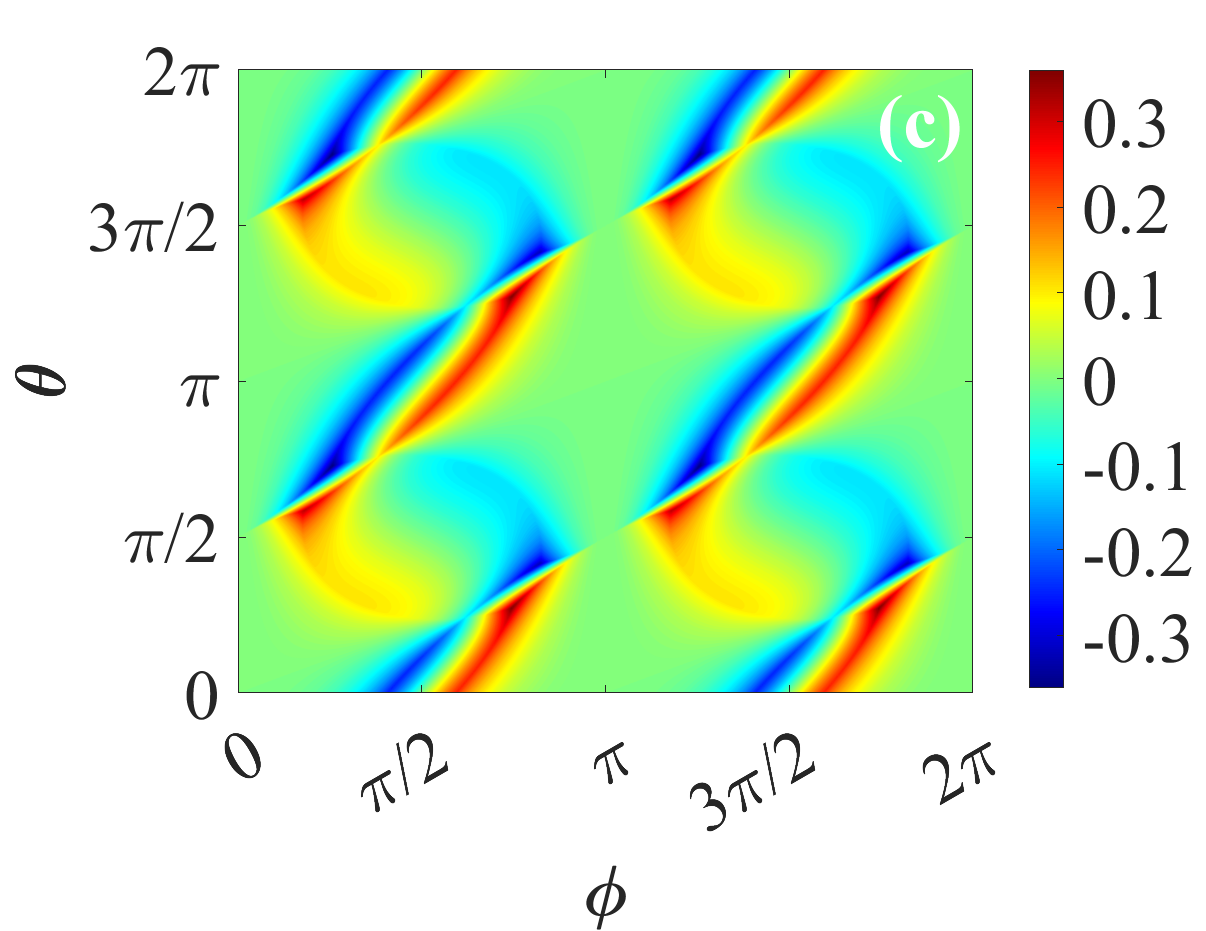}
\caption {Diode rectification amplitude of the Josephson junction between a single-band and a three-band superconductor as a function of phases differences in a three-band superconductor for (a) $R_{N1}=R_{N2}=R_{N3}$, (b) $R_{N1}/R_{N2}=1$, $R_{N1}/R_{N3}=2$ and (c) $R_{N1}/R_{N2}=2$, $R_{N1}/R_{N3}=3$ at zero temperature.}
\label{Diod_3band}
\end{figure}

To avoid cumbersome expressions we further assume the coincidence of the order parameter amplitude in superconducting leads: $\left| {{\Delta _0}} \right| = \left| {{\Delta _1}} \right| = \left| {{\Delta _2}} \right| = \left| {{\Delta _3}} \right|$. Moreover, bearing in mind that in the case under consideration we deal with the exact solutions of the Eqs. (\ref{Usadel_2band}) for Green's functions, we can treat the case at zero temperature by summing over Matsubara frequencies in Eq. (\ref{current_total}). The resulting Josephson current is given by the following expression:
\begin{eqnarray}
\label{CPR_3band_T=0}
I\left( \varphi  \right) &=& \frac{{\pi \left| \Delta_0  \right|}}{{e{R_{N1}}}}\cos \frac{\varphi }{2}{\mathop{\rm Artanh}\nolimits} [\sin \frac{\varphi }{2}] \nonumber \\
 &+& \frac{{\pi \left| \Delta_0  \right|}}{{e{R_{N2}}}}\cos \left( \!{\frac{\varphi }{2} + \phi } \!\right){\mathop{\rm Artanh}\nolimits} [\sin \left(\! {\frac{\varphi }{2} + \phi }\! \right)]\nonumber \\
 &+& \! \frac{{\pi \left| \Delta_0  \right|}}{{e{R_{N3}}}}\cos \left(\! {\frac{\varphi }{2} + \theta }\! \right){\mathop{\rm Artanh}\nolimits} [\sin \left(\! {\frac{\varphi }{2} + \theta } \!\right)].
\end{eqnarray}

Eq. (\ref{CPR_3band_T=0}) can be regarded as a generalization of the classical formula, derived by Kulik and Omelyanchouk for the S-c-S contact at T=0 \cite{Kulik1, Kulik2}, for the case of the Josephson junction between the conventional s-wave and three-band superconductor. As expected (Figure \ref{CPR_3band}a) at $\phi=0$ and $\theta=0$ and the critical currents turn out to be reciprocal: $I_c^+=|I_c^-|$.

Instead, for the three-band superconductor with a time reversal broken order parameter (when $\phi$ and $\theta$ are different from zero and/or $\pi$), the current phase relation exhibits a complex non-sinusoidal pattern (Figure \ref{CPR_3band}b). Indeed, one can observe that for certain values of the intrinsic phase differences $ \{\phi,\theta\}$ a strong asymmetry of critical currents is achieved (Fig. \ref{CPR_3band}b).
There, the profile of the current-phase relation presented in Fig. \ref{CPR_3band}b can provide 
a sizable rectification value $\eta \approx 0.25$.

The overall rectification amplitude of the Josephson junction with three-band superconducting lead is reported in Fig. \ref{Diod_3band} for different cases of band-dependent tunneling amplitude by scanning the whole space of interband phase relation. 
In Fig. \ref{Diod_3band}a the symmetric case with all the bands having the same resistance is shown.
The profile of the rectification can be understood by inspection of the harmonic content of the current phase relation in Eq. \ref{CPR_3band_T=0}.
Indeed, at $\phi=\pi/3$ and $\theta=\pi/6+(2 n+1) \frac{\pi}{2}$ (for $n=1,2,..)$ all the even harmonic components are vanishing. Hence, the rectification amplitude is zero because the current phase relation has a fixed parity. Since the parity of the current phase relation cannot be fixed in isolated points of the phase space there must be nodal lines. 
Indeed, along the lines defined by the relation $\theta=(2 n+1) \pi/2+\phi$, $\theta=(2 n+1) \pi/2+\phi$, $\theta= n \pi-\phi$, the even harmonics of the second and third terms of Eq. \ref{CPR_3band_T=0} cancel out. This implies that the rectification amplitude has to vanish. 
Since the rectification amplitude can be maximized when the first and the second harmonics components have comparable amplitude we expect that moving away from the nodal lines there will be regions where the rectification increases. This is indeed a feature of the pattern of the rectification amplitude (Fig. \ref{Diod_3band}a) with phase space domains of sizable rectification strength developing around the nodal lines. 
A modification of the tunneling amplitude $R_{N}$ alters the balance between the terms which are dependent on $\theta$ and $\phi$ leading to a variation of the nodal lines (Fig. \ref{Diod_3band}a) and of the rectification pattern around them. 

Finally, we point out that for the diagram in Figure \ref{Diod_3band},
the states close to $\theta=0$ and $\phi=0$ lines do not fulfill the stability conditions and are metastable configurations \cite{Brink}. 
The same arguments apply for the states  $\{\phi\!=\!0, \theta\!=\!0 \!\} $, $\{\phi\!=\!0, \theta\!=\!\pi \!\} $, $\{\phi\!=\!\pi, \theta\!=\!0 \} $, $\{\phi\!=\!\pi, \theta\!=\!\pi \} $. For other points of the phase diagram is instead possible to set the intra- and interband interaction (elements of 3-by-3 matrix), which satisfy the conditions of achieving a stable ground state.

As a final remark, we would like to note that this model can be generalized to the case of a larger number of order parameters in a multiband superconductor. We do not present here the diagram of a superconducting diode for the case of a pure $n$-band superconductor because of technical difficulties in its visualization. However, it is obvious that in this case the rectification amplitude pattern becomes even more nontrivial. 

\section{Conclusions}

The effects of nonreciprocal supercurrent in Josephson junctions have been studied using multiband superconductors that can disrupt time-reversal symmetry by interband phase reconstruction. It has been shown that nonreciprocal supercurrent can be achieved through a combination of interband superconducting phase mismatch and scattering, as well as by exploiting multiband phase frustration. The presence of impurity scattering between bands affects the magnitude and direction of the nonreciprocal supercurrent, depending on the interband phase relationship. In the case of a three-band superconductor with a phase-frustrated configuration, the supercurrent rectification profile is characterized by a hexagonal pattern of nodal lines with zero amplitude. 
Interestingly, the amplitude of supercurrent rectification shows a three-fold pattern with alternating signs around the nodal lines. We have found that the hexagonal pattern and the three-fold structure in the interband phase space are influenced by the band dependent tunneling amplitudes. These findings applied to corresponding diodes can be used to detect a state with the time-reversal symmetry breaking in multiband superconductors due to a nontrivial interband phase relation.

Finally, we point out that the behavior of the junction with three-band superconductor interfaced to a one-band superconductor can be also mimicked by designing a double loop superconducting quantum interference device. Indeed, recently a double loop interferometer based on conventional superconductor-normal-superconductor weak links has been demonstrated to yield a rectification hosting nodal lines and multifold sign tunable patterns \cite{Greco2024}. We argue that the use of such double loop interference device based on multiband superconductors can be employed to disentangle the interband phase complexity.

\section*{Acknowledgments}
M.C. and F.G. acknowledge financial support from the EU’s Horizon 2020 Research and Innovation Framework Program under Grant Agreement No. 964398 (SUPERGATE) and from PNRR MUR project PE0000023-NQSTI. M.C. acknowledges support from the QUANCOM Project 225521 (MUR PON Ricerca e Innovazione No. 2014–2020 ARS01 00734). F.G. acknowledges the EU’s Horizon 2020 Research and Innovation Framework Programme under Grant No. 101057977 (SPECTRUM) for partial financial support.

\bibliography{REF}

\end{document}